# Imaging the source of high-harmonics generated in atomic gas media

**STEFANOS CHATZIATHANASIOU,**[1,3] **SUBHENDU KAHALY,**[2] **DIMITRIS CHARALAMBIDIS,**[1,2,3] **PARASKEVAS TZALLAS,**[1,2] AND **EMMANOUIL SKANTZAKIS**[1*]

[1]*Foundation for Research and Technology-Hellas, Institute of Electronic Structure & Laser, P.O. Box 1527, GR-71110 Heraklion (Crete), Greece*
[2]*ELI-ALPS, ELI-Hu Kft., Dugonics ter 13, 6720 Szeged, Hungary.*
[3]*Department of Physics, University of Crete, PO Box 2208, GR71003 Heraklion (Crete), Greece*
\* *skanman@iesl.forth.gr*

**Abstract:** We report the application of the time gated ion microscopy technique in accessing online the position of the source of harmonics generated in atomic gas media. This is achieved by mapping the spatial extreme-ultraviolet (XUV)-intensity distribution of the harmonic source onto a spatial ion distribution, produced in a separate focal volume of the generated XUV beam through single photon ionization of atoms. It is found that the position of the harmonic source depends on the relative position of the harmonic generation gas medium and the focus of the driving infrared (IR) beam. In particular, by translating the gas medium with respect to the IR beam focus different "virtual" source positions are obtained online. Access to such online source positioning allows better control and provides increased possibilities in experiments where selection of electron trajectory is important. The present study gives also access to quantitative information which is connected to the divergence, the coherence properties and the photon flux of the harmonics. Finally, it constitutes a precise direct method for providing complementary experimental info to different attosecond metrology techniques.

## 1. Introduction

Time gated ion microscopy [1-4] has been successfully employed in recent years for (a) the in situ focus diagnostics [1,2,4], (b) quantitative studies of linear and non-linear ionization processes both in the infrared (IR) and the extreme ultraviolet (XUV) regimes [1,4], and proposed for single-shot XUV-pump-XUV-probe studies as well as for single-shot 2nd-order XUV autocorrelation measurements [4,5]. The approach is based on the measurement of a spatial ion distribution resulting from the interaction of the radiation with gas phase media. High Harmonic Generation (HHG) [6] results routinely from the interaction of intense focused fs IR laser pulses with gases, solid targets and recently with nanostructures [7-12]. In atomic gases, it is well understood that in the strong field regime there are two electron trajectories within each laser half cycle, called the *long (L)* and the *short (S)*, with different excursion times in the continuum which contribute to emission at each harmonic frequency. It turns out that the different continuum excursion times manifest in different phase contributions to the *L* and *S* electron trajectories leading to distinctive features. Harmonic radiation which is generated mainly by the *L* trajectories present higher divergence than this generated mainly from the *S* one [13-15]. Furthermore, their relative contribution to the outgoing from the medium harmonic beam can be controlled by the appropriate focusing geometry [14]. In spite of the knowledge acquired over the last two decades on the role of the electron trajectories on the spatial properties of the generated harmonic beam, in the majority of the experiments it is considered that the image of the XUV source is unaffected by the geometrical conditions in the HHG area. In particular, the influence of the relative position of the IR focus and the gas jet on the "virtual" position of the harmonic source has never been directly demonstrated experimentally. Thus, this work constitutes a simple, direct manifestation of a phenomenon underlying previous works [13-17]. The aim of the present work is to study the dependence of the "virtual" position of the harmonic source on the laser focusing geometry using an Ion-Microscope (I-M) imaging detector. By "virtual" position here we mean the position at which the source has to be considered being placed in order to produce the measured image after refocusing of the XUV beam. I-M allows the recording of the spatial distribution of the ionization products produced by a (usually focused) beam and consequently the spatial intensity distribution of the ionizing XUV radiation. When ionization occurs at the focus of the XUV radiation, the spatial ion distribution is an image of the radiation source itself, which in our case is the area where the XUV beam is generated. The operational principle of I-M can be found elsewhere [4]. Taking advantage of the high spatial resolution of the I-M we have been able to study the variation of the XUV focus position as a function of the relative position of the gas medium and the IR focus. This is realized by varying the position of the gas medium and recording the spatial ion distribution of Ar produced by a single-XUV-photon ionization process at the focus of the XUV beam keeping the position of the IR focus constant. The present study allows the precise direct measurement of the XUV beam divergence generated in gas phase media. In addition, it can provide quantitative information during the coherent synthesis of sequentially positioned high harmonic sources when different phase matching and quasi-phase matching approaches are employed [18]. These approaches are directly connected to the improvement of the coherence properties and the photon flux of the attosecond pulses. Finally, it gives access to supplementary experimental info which is linked to different, recently proposed and implemented, attosecond metrology techniques like [19-22].

## 2. Experimental set-up

The experiment is performed utilizing a 10 Hz Ti:Sapphire laser system delivering 25 fs long laser pulses with central wavelength 800 nm and energy ≈15 mJ/pulse. The experimental set up is shown in Fig. 1a. The laser beam is focused with a f = 3m lens (L) into a pulsed gas jet (P-GJ) with a rectangular orifice with dimensions 0.3 mm x 1 mm, filled with Ar gas. The gas jet is mounted on a translation stage allowing the variation of the position of the jet with respect to the IR focus (see grey arrow in Fig. 1(a)). The IR radiation is eliminated after the XUV generation by a combination of a reflection on a silicon plate (Si) placed at the Brewster angle of the IR beam, a 5-mm diameter aperture (A) and a 150 nm thick Sn filter which transmits the harmonics (q) from 11th to 15th. Subsequently, the XUV radiation is focused into the target gas jet (T-GJ) filled with Ar by a spherical gold mirror (SM) of 5 cm focal length. Care has been taken to fix the angle of incidence of the XUV beam on the gold mirror at ≈ 0°. The SM is placed 4.5 m downstream the P-GJ. The images are monitored by the transversely placed I-M which records the spatial distribution of Ar ions as described above (Fig. 1(b)).

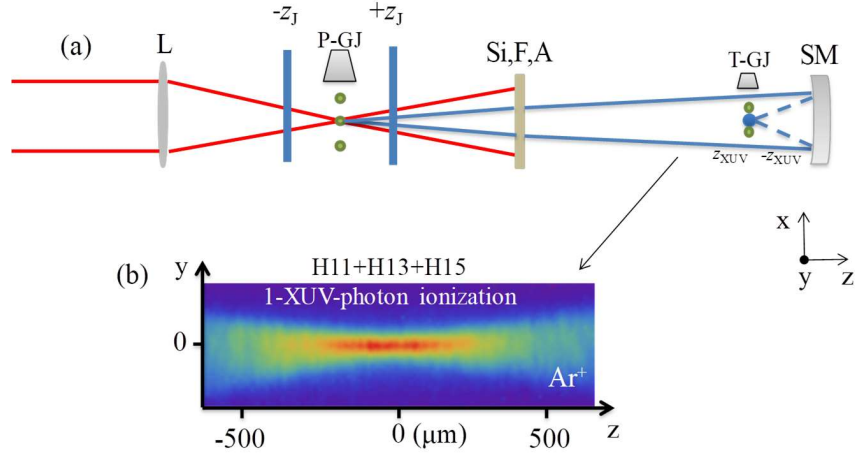

Fig. 1. (a) Schematic of the experimental set-up. L: Lens P-GJ: Pulsed gas jet used for the HHG mounted on translation stage of micrometer precision; Si: Silicon plate, A: Aperture, F: Filter, T-GJ: Target gas jet, SM: Spherical mirror (b) A typical image of Ar$^+$ spatial ion distribution produced by a single-XUV-photon ionization process at the focus of the XUV beam, as it was retrieved by the I-M after the accumulation of 600 shots. The XUV beam in the T-GJ area consists the harmonics from 11th to 15th with relative amplitudes 0.6(11th):1(13th):0.8(15th).

## 3. Results and discussion

We vary the position ($z_j$) of the pulse gas jet P-GJ from $z_j \approx$ -3 cm to $z_j \approx$ +3 cm with a step of 3 mm while recording the XUV focus distribution (in x-z plane in Fig. 2(a)) and location (along z) with respect to the P-GJ position corresponding to each $z_j$. The range of the scan of the P-GJ is almost coinciding with the full confocal parameter b≈7 cm of the focused driver. In the single XUV photon ionization limit, the HHG yield is proportional to the total ion yield measured by I-M. Initially, by spatially integrating the recorded images we have obtained the dependence of HHG yield on the position of the P-GJ (Fig. 2 (a)). The dependence of the XUV focus position ($z_{XUV}$) on the position of P-GJ is shown in Fig. 2(b). The position of the XUV focus was defined by the maximum of the ion distribution obtained from the lineout at y=0 along the propagation axis (z) (Fig. 1(b)). The maximum was obtained by the zero of the derivative of the line outs. In order to avoid the noise introduced in the derivative by the fine structure (of few pixel size) of the ion distribution, the line outs have been smoothed over 70 points. The overall displacement of the $z_{XUV}$ found to be ≈ 20 μm and cannot be explained by the 6 cm movement of P-GJ (which can provide a maximum displacement of ≈ 7.6 μm at the XUV focus). Considering that i) for a specific position of the P-GJ, ii) the non-linearity of the HHG process results to a HHG yield $\propto (I_L)^p$ where $I_L$ is the intensity of the laser and $p$ = 3-5 for all plateau harmonics [23] and assuming that all the harmonics are emitted from the same cross section in the gas medium, the observed displacement is attributed to the different wave front curvature, which for Gaussian beams essentially means to the different divergence of the XUV beams resulting from the $L$ and $S$ trajectories. Particularly it was shown [24] that the dependence of the divergence on the electron trajectories for the harmonic order $q$ is given by

$\theta_{S,L} = \frac{\lambda_q}{\pi w_q}\sqrt{1 + 4\alpha_{S,L}^2 I_L^2 \frac{w_q^4}{w_f^4}}$ (eq.1) (where $\lambda_q$, $w_q$, are the wavelength and the beam waist of the harmonic $q$, $w_f$ is the waist size of the laser beam and $\alpha_{S,L}$ is the S, L trajectory coefficient [24]), which results in $\theta_S/\theta_L = \sqrt{(1+A\alpha_S^2)/(1+A\alpha_L^2)} \approx \alpha_S/\alpha_L$ (eq.2). Although the contribution of both electron trajectories at each P-GJ position cannot be excluded [25], in the present study we can safely consider that for $z_j \approx$ +3 cm (z $_j\approx$ b/2, where $b$ is the confocal parameter and b≈7 cm) and $z_j \approx -3$ cm the main contribution is coming from the *S* and *L* trajectory harmonics, respectively. Differences that phase matching imposes to the radiation emitted by the two different quantum trajectories such as the different divergence of them, have already been established before. They have been extensively studied in the past [13,14] and are directly verified in the present work by measuring the width of the line out along the propagation axis for each position of the P-GJ (Fig. 2 (c)). It is evident that for $z_j \approx 0$ where both trajectories have significant contribution to the harmonic emission [2], the width of the distribution along the propagation axis is ≈ 1.5 times larger than the width measured at $z_j \approx \pm$ 3cm where the main contribution is coming from single trajectory harmonic.

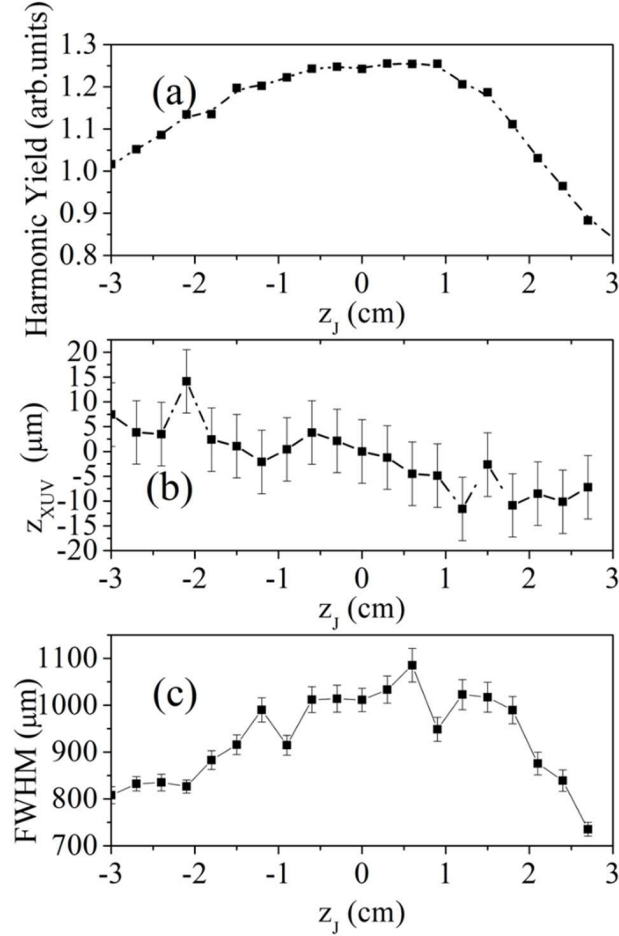

Fig. 2. (a) Spatially integrated Ar$^+$ ion signal produced by a single-XUV-photon ionization process at the focus of the XUV beam as a function of the gas jet position (b) XUV focus position as a function of the gas jet position. The error bars represent one standard deviation of the mean value. (c) FWHM of the XUV focal distributions resulting from a gaussian fit as a function of the gas jet position. The error bars correspond to the error of the FWHM after the gaussian fit of the lineouts. 600 shots were accumulated for each image.

Thus, considering as central wavelength of the XUV the 13th harmonic ($\lambda_h$= 61.5 nm), $\alpha_S \approx 2.5 \times 10^{-14}$ rad cm$^2$/W, $\alpha_L \approx 22.5 \times 10^{-14}$ rad cm$^2$/W [26], and $w_f = w_q \approx 150 \mu m$ [24], the above relation results to $\theta_S/\theta_L \approx 0.12$. Utilizing geometrical optics, this ratio leads to a distance of $\Delta z_{obj} \approx 22$ cm between the "virtual" positions of the *S*, *L* trajectory-harmonics sources and results to an overall displacement of $\Delta z_{XUV} \approx 23$ µm at the focus of the XUV. Incorporating the displacement introduced in P-GJ this value is in excellent agreement with the experimental findings. Fig. 3 shows a schematic which depicts the "virtual" positions of the harmonic source ($V_S$ and $V_L$) and the corresponding XUV foci at $z_j = -3$ and $z_j = 3$ ($Im_L$ and $Im_S$, respectively).

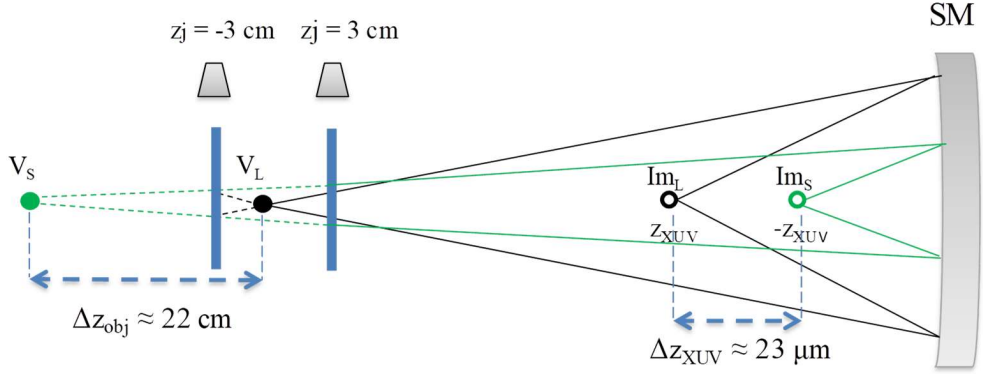

Fig. 3. By varying the generation jet position from $z_j \approx -3$ cm to $z_j \approx 3$ cm "virtual" positions of the harmonic source are resulted due to the varying divergence difference of the harmonic radiation emitted by the *S* and *L* trajectories (blue vertical line). At the jet position $z_j = -3$ cm the *L* trajectories with higher divergence (black dashed lines) dominate the harmonic emission leading to a "virtual" positions of the harmonic source $V_L$. On the contrary, at the jet position $z_j = 3$ cm the *S* trajectories with smaller divergence (green dashed lines) dominate the emitted radiation leading to a "virtual" position of the harmonic source $V_S$. The difference of the position of the "virtual" sources is $\Delta z_{obj} \approx 22$ cm. This difference in the "virtual" positions should lead to a $\Delta z_{XUV} \approx 23$ µm displacement of the XUV focus position ($Im_L$ and $Im_S$) after the reflection on the spherical mirror which is in accordance with the geometrical optics and in fairly good agreement with the measured experimental value of $\approx 20$ µm.

Generally, the divergence changes with harmonic order. In this work we measure the average effect of all three not resolved harmonics involved (11, 13, 15) and the calculations are done for the central one i.e. the 13$^{th}$ harmonic. This is justifiable because applying (eq. 1) to all three harmonics one can easily extract that the ratio $\theta_L/\theta_S$ is essentially the same for all three harmonics, namely, 0.112, 0.113 and 0.111 for the 11$^{th}$, 13$^{th}$ and 15$^{th}$ harmonic respectively. While the atomic dipole phase is an unambiguous factor contributing to the effect at hand other propagation effects such as defocusing due to ionization and/or third order non-linearities [27,28] are much less probable, as I) the intensity was safely kept below saturation, II) the medium we use is very short compared to the confocal parameter and III) such phenomena is expected to have the same effect on both trajectories. Nevertheless, our experimental approach can be used also for the study of such phenomena whenever they become relevant

## 4. Conclusions

In conclusion, by utilizing a time gated ion microscopy technique, we demonstrate that the image of the XUV source is affected by the geometrical conditions in the HHG area. Particularly, it was found that the short- and long-trajectory harmonics generated in gas phase media are focused in different positions in the detection area. This was measured by recording the dependence of spatial ion distribution produced by single-XUV-photon ionization process at the XUV focus on the position the HHG medium

## 5. Funding

We acknowledge support of this work by: the LASERLAB-EUROPE (grant agreement no. 284464, EC's Seventh Framework Programme), H2020-INFRAIA project NFFA-Europe (Nr.: 654360), "HELLAS-CH" (MIS 5002735) (which is implemented under the "Action for Strengthening Research and Innovation Infrastructures", funded by the Operational Programme "Competitiveness, Entrepreneurship and Innovation" (NSRF 2014-2020) and co-financed by Greece and the European Union (European Regional Development Fund)) and the European Union's Horizon 2020 research and innovation program under Marie Sklodowska-Curie grant agreement no. 641789 MEDEA. ELI-ALPS is supported by the European Union and co-financed by the European Regional Development Fund (GINOP-2.3.6-15-2015-00001).

## 6. Acknowledgments

We thank N. Papadakis for developing the electronic devices and data acquisition software used in the beam line.